\documentclass{article}
\usepackage{amsmath}
\usepackage{amssymb}
\usepackage{cite}
\usepackage {hyperref}
\usepackage[dvips]{graphicx}

\numberwithin{equation}{section} 

\begin{document}

\title{Chaos, Fractals and Quantum Poincar\'e Recurrences in
Diamagnetic Kepler Problem}
\author{{A. Ugulava, L. Chotorlishvili*, T. Kereselidze, V.
Skrinnikov} \\
{\small\textit{Department of Physics, Tbilisi State University,}} \\
{\small\textit{Chavchavadze av. 3, 0128 Tbilisi, Georgia}} \\
{\small\textit{*E-mail: lchotor33@yahoo.com}}}
\date{}
\maketitle

\begin{abstract} The statistics of quantum
Poincar\'e recurrences in Hilbert space for diamagnetic hydrogen
atom in strong magnetic field has been investigated. It has been
shown that quantities characterizing classical chaos are in a good
agreement with the ones that are used to describe quantum chaos.
The equality of classical and quantum Poincar\'e recurrences has
been shown. It has been proved that one of the signs of the
emergence of quantum chaos is the irreversible transition from a
pure quantum mechanical state to the mixed one.
\end{abstract}


\section{Introduction}

The investigation of the chaotic processes in the deterministic
systems is one of the most important trends of the modern physics
\cite{1,2,3,4,5}. The successes of this current for nonlinear
classical systems that are described via trajectories of the
system in Hilbert space are obvious \cite{6,7,8}. Yet the
conception of deterministic chaos in quantum systems, where the
idea of trajectory is inapplicable, stays a subject for serious
discussions so far. There is no common understanding of the
problem of quantum chaos for today. Usually the quantum chaos is
considered as a set of phenomena taking place in quantum systems,
the classical analogues of which display chaos
\cite{9,10,11,12,13,14,15,16}.

Generally when investigating the problems of quantum chaos, one
has to deal with the Hamiltonians of the form:
\begin{equation}
\hat{H}(p,q,\lambda)=\hat{H}_{0}(p,q)+\lambda V(q),
\end{equation}
where p, q are a set of classical coordinates and momentum, $
\hat{H}_{0}(p,q) $ is classically integrable part of Hamiltonian,
$ V(q) $ is a part of Hamiltonian which leads to the
nonintegrability of the classical equations of motion when added,
$ \lambda $ is a parameter by variation of which the system may be
driven in the domain of chaotic dynamics.

The goal of this work is the investigation of a concrete physical
system described by the Hamiltonian of the form (1.1). The object
of our investigation is a diamagnetic hydrogen atom placed in a
strong magnetic field. Diamagnetic hydrogen atom, subjected to the
influence of the strong magnetic field, has been considered in the
quasi-classical approximation in the ref \cite{16}. Namely a
photo-absorbtion spectrum in the transition regime to chaos has
been studied. The main purpose of our work is the investigation of
the irreversible evolution of the quantum-mechanical system and
the passage from the quantum-mechanical to the quantum-statistical
consideration. We shall try to compare and find out the link
between the phenomena taking place under both classical and
quantum considerations.

Relatively weak magnetic field applied to atom leads just to the
energy levels splitting, i.e. to the Zeeman effect. In case when
the hydrogen atom is placed in the extreme high magnetic field,
the motion of the electron in the plane perpendicular to the field
direction is completely defined by this magnetic field and not by
the coulomb field of the nucleus, and because of this, the atom is
found to be deformed in the transverse direction. At the same time
the field does not affect on the lengthwise motion, and in this
direction the atom keeps its size.

One may easily estimate the value of the magnetizing force when
the deformation of the electron shell takes place. One has to
compare the value of Bohr radius $a_0=\hbar^2/m_ee^2$ (where
$\hbar$ is a Plank constant, $m_e$ and $e$ are the electron mass
and charge respectively) with a characteristic size of the
localization domain of electrons in the magnetic field in the
ground state when $n=0$, $s=0$, $a_B=\sqrt{\hbar c/eB}$. If
$a_B<a_0$, then the magnetic field has the defining effect. This
condition leads to the following estimate of the extremely strong
field
\begin{equation}
B>\frac{m_e^2ce^3}{\hbar^3}=B_c=2.35\cdot10^5 Ts,
\end{equation}
in which the atom is deformed. Notice, that the magnetic fields
required for this, the intensity of which is defined by the
condition (1.2), according to the modern concepts, may exist on
the surface of some astrophysical objects. The neutron stars (or
pulsars) appearing as the result of the collapse of the supernova,
belong to them. That is why the investigation of the diamagnetic
atoms in the conditions of extremely high magnetic fields is of
doubtless interest.

\section{Classical Consideration}

Consider the hydrogen atom placed in the magnetic field directed
along z-axis $\vec{B}=B\vec{e}_z$. We shall use Larmor theorem.
According to this theorem, for studying the diamagnetism theory
for single atoms it is enough to consider the motion of the
electron in the coordinate system, rotating with the Larmor
frequency $\vec{\omega}_L=\gamma_e\vec{B}$, where $\gamma_c$ is a
gyromagnetic ratio for electrons.

During the transition to the rotating coordinate system, the
change of the kinetic energy of the electron will occur. This
change is conditioned by the addition to the velocity of in the
rotating coordinate frame $\vec{u}$ the cross-product of the
Larmor frequency $\vec{\omega}_L$ and the vector $\vec{r}$ of the
location of electron. Taking into account, that the direction of
vector $\vec{\omega}_L$ is opposite to the vector
$\gamma_e\vec{B}$, and gyromagnetic ratio of the electron is
$\gamma_e=-\frac{e}{2m_ec}$ (here $c$ is the velocity of light),
we shall achieve for the velocity of electron in a motionless
coordinate frame:
\begin{equation*}
\vec{v}=\vec{u}+[\vec{\omega}_L\vec{r}]=\vec{u}+\frac{e}{2m_ec}[\vec{B}\vec{r}].
\end{equation*}
Then the kinetic energy of the electron will be of form:
\begin{equation*}
\begin{split}
K=\frac{1}{2}m_eu^2+m_e[\vec{\omega}_L\vec{r}]\vec{u}+\frac{1}{2}m_e\lvert[\vec{\omega}_L\vec{r}]\rvert^2=\\
=\frac{1}{2}m_eu^2+\frac{e}{2m_ec}BL_z+\frac{e^2}{8m_ec^2}B^2(x^2+y^2).
\end{split}
\end{equation*}

One may easily see, that if z-projection of the angular momentum
is equal to zero, $L_z=0$ the task can be reduced to the
diamagnetic Kepler problem. Finally in the system of atomic units
the Hamiltonian of the considered problem takes the form:
\begin{equation}
H=\frac{1}{2}p^2-\frac{1}{r}+\frac{1}{8}\lambda(x^2+y^2)=E,
\end{equation}
where $\lambda=\frac{\hbar^3}{m_e^2ce^3}B$ is a constant
describing the connection of the system with the external magnetic
field. The form of the Hamiltonian (2.1) is similar to the one of
(1.1). For $\lambda=0$ the equations corresponding to (2.1) are
integrable. For $\lambda\neq0$ the system is nonintegrable.
Finally with the increase of the amplitude of the magnetic field,
starting with some definite value of $\lambda_c$, the dynamics
appears to be chaotic. The use of the numerical methods for the
Hamiltonian of the form (2.1) is connected with definite
difficulties because of the appearance of singularity in the $r=0$
point.

To get over this problem it is necessary to use semiparabolic
coordinates that are connected with the Cartesian ones via
formulas:
\begin{equation*}
\begin{cases}
\mu=\sqrt{r+z}\\
\nu=\sqrt{r-z}
\end{cases}
\end{equation*}
and make a transition to a non-homogeneous time via
transformation $dt=(\nu^2(\tau)+\mu^2(\tau))d\tau$ \cite{15}. Then
taking into account that $L_z=0$, the Hamiltonian (2.1) can be
transformed to the following form \cite{15}:
\begin{equation}
\frac{1}{2}(p_\mu^2+p_\nu^2)-E(\mu^2+\nu^2)+\frac{1}{8}\lambda\mu^2\nu^2(\mu^2+\nu^2)=2.
\end{equation}
I.e. the Hamiltonian is of form
\begin{equation}
H(p_\mu,p_\nu,\mu,\nu,\lambda)=H_0(p_\mu,p_\nu,\mu,\nu)+\lambda
V(\mu,\nu),\tag{2.2a}
\end{equation}
where $H_0=\frac{1}{2}(p_\mu^2+p_\nu^2)-E(\mu^2+\nu^2)$ is the
integrable part of the Hamiltonian;
$V=\frac{1}{8}\mu^2\nu^2(\mu^2+\nu^2)$ is the nonintegrable part
of the Hamiltonian.

As one may see from (2.2a), the transformed Hamiltonian by its
form coincides with the Hamiltonian of interacting oscillators,
and in (2.2) the singularities which are typical for the
Hamiltonian (2.1) are absent. Though one should take into account,
that during the direct numerical calculations, right after the
integration of the Hamilton equations corresponding to (2.2a), one
has to perform a reverse transition to a real time.

Henceforth the transformed Hamiltonian will be the object of our
investigations in case of classical consideration.

As it is clear from (2.2), the value of the magnetic field $B$ is
the parameter of nonlinearity. We shall study a case of a strong
nonlinearity, i.e. the case when the methods of perturbation
theory are inapplicable. In such case the chaos may appear in the
system. We should expect to get a chaos in case of high magnetic
fields $B\ge B_c$.

When chaos appears, the mechanical concepts lose their meaning.
For describing the processes taking place in the system, the
statistical concepts: Kolmogorov entropy \cite{7} and the fractal
dimension of the phase space of the system \cite{17,18,19} become
significant. The study of the statistical features of the
dynamical system can be performed only numerically.

The results of the numerical computations, gained by solving the
Hamilton equations, corresponding to (2.2), are given in Fig.1 and
Fig.2.

\begin{figure}[tbp] \centering
\includegraphics[width=7.5cm]{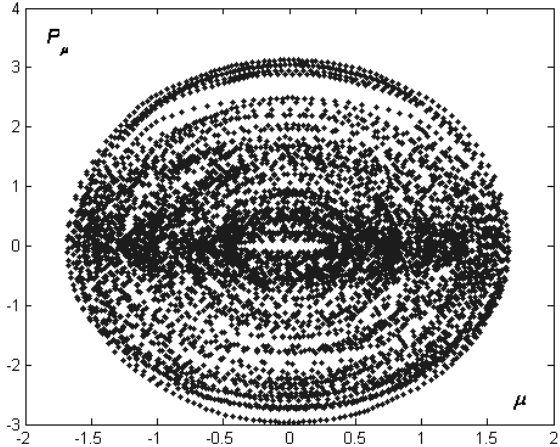}

\caption{\small Phase trajectory on the plane of variables
$(p_\mu,\mu)$, obtained after the numerical integration of the
Hamilton equations, corresponding to the Hamiltonian (2.2). The
result (as all the other numerical calculations given in this
work) is achieved by using the software MathWorks MatLab 6.0 for
the values of parameters $E=-0.125\sim10^{-19} J$, $\lambda=1.0$.
As it is seen from the graph, the motion is of a chaotic type, and
the destruction of the invariant tori points to
this.}\label{Fig:1}
\end{figure}

\begin{figure}[tbp]
\centering
\includegraphics[width=7.5cm]{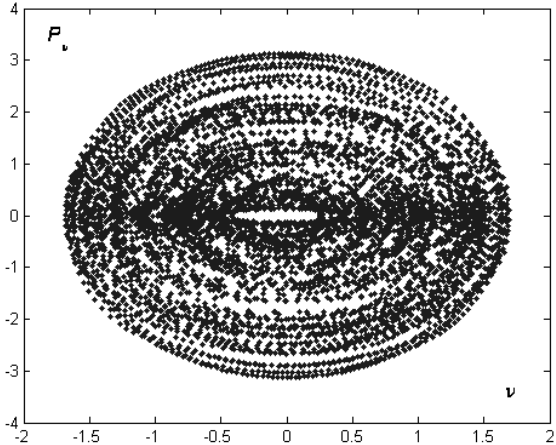}
\caption{\small Phase trajectory on the plane of variables
$(p_\nu,\nu)$ plotted under the conditions similar to the one in
Fig.1. The destructed invariant torus, according to KAM theorem
\cite{7}, testifies to the existence of the classical
chaos.}\label{Fig:2}
\end{figure}

As seen from Fig.1 and Fig.2, the destruction of the invariant
tori took place. The motion is of the chaotic type. For more
convincing proof of the chaotic character of motion one has to
perform Fourier analysis of the obtained numerical data.

To define Kolmogorov entropy one has to perform a Fourier
transformation of the correlation function
\begin{equation}
\begin{split}
G_\mu(\tau)&=\langle\mu(t+\tau)\mu(t)\rangle,\\
G_\nu(\tau)&=\langle\nu(t+\tau)\nu(t)\rangle,\\
G(\omega)&=\int G(\tau)e^{\imath\omega
\tau}d\tau=\frac{\tau_c}{1+\omega^2\tau_c^2},
\end{split}
\end{equation}
where
\begin{equation*}
\langle(\dotsb)\rangle=\lim_{T\rightarrow\infty}\frac{1}{T}\int_0^T(\dotsb)dt,
\end{equation*}
means averaging over the time, $\tau_c$ is the correlation time,
connected with the Kolmogorov entropy by the ratio
$h_0\sim\frac{1}{\tau_c}$. In practical calculations of (2.3) it
is more convenient to use a realization of a Fast Fourier
Transform algorithm \cite{20,21}. As the numerical calculations
show, the value of the correlation time is $\tau_c\sim10^{-15} s$.
The results of numerical calculations are presented in Fig.3.
\begin{figure}[tbp]
\centering
\includegraphics[width=7.5cm]{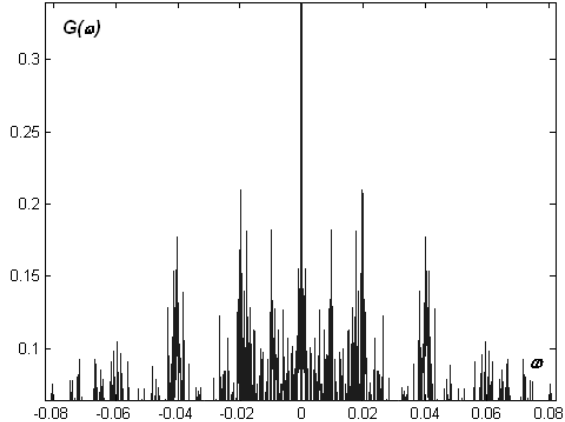}
\caption{\small The dependence of the Fourier image of the
correlation function $G_\mu(\omega)$ on the frequency $\omega$
($10^{15}Hz$), plotted using Eq. (2.3) and FFT method. The
presence of the finite width of the correlation function
$G_\mu(\omega)$ is the evidence of the appearance of chaos. In other
words, one may say, that oscillatory process is not characterized
by a definite frequency (period). All the frequencies (periods) in
a finite frequency interval $\delta\omega\sim1/\tau_c=200 THz$ are
involved into oscillatory process.}\label{Fig:3}
\end{figure}

There is one more interesting property of dynamical systems
connected with the Kolmogorov entropy that describes chaotic
motion and is called Poincar\'e recurrence. The main point of the
Poincar\'e recurrence is that any system with a finite phase space
after some definite time returns to its initial state \cite{22}.
The study of the time of Poincar\'e recurrence is one of the most
powerful methods of the analysis of the nonlinear systems.

It turned out to be that in conditions of strong chaos (i.e. when
the phase plane doesn't contain the stability islands) the
distribution of the times of Poincar\'e recurrences has a Lorenz
structure \cite{22}:
\begin{equation}
P(\tau)=\frac{1}{\tau_{rec}}\exp(-\tau/\tau_{rec}),
\end{equation}
where $\tau_{rec}=\int_0^\infty \tau P(\tau)d\tau$ is the average
recurrence time connected with Kolmogorov entropy as
$\tau_{rec}=\frac{1}{h_0}$. Thus, calculating the value of the
Kolmogorov entropy one can define the distribution of the
recurrence times.

We in details touched upon the question about classical Poincar\'e
recurrences, because henceforth we will be interested in quantum
Poincar\'e recurrences. The idea of quantum Poincar\'e recurrence
was introduced in the recently published work by G. M. Zaslavsky
and A. Iomin \cite{22}. We shall return to the description of this
phenomenon during a quantum consideration.

One of the signs of the appearance of chaos is a discrete
(fractal) dimension of the phase space of the system.

For estimation of the fractal dimension we shall use the
Grassberger-Procaccia algorithm \cite{17,18,19}. The main point of
this algorithm consists into following:

Assume we have obtained from a numerical solution of the equations
of motion a set of state vectors $\lbrace x_i,
i=1,2,\dotsb,N\rbrace$ that correspond to the sequential steps of
integration. In our case $x_i$ denote a complete set of variables,
characterizing the phase space of the system: $P_\mu(t_i)$,
$P_\nu(t_i)$, $\mu(t_i)$, $\nu(t_i)$, $t_1=0$, $t_N=T$, where
$t_i\in[0,T]$ is the time interval for the numerical integration.
Defining some (small) $\epsilon$, we can use our set
for estimation of the following sum:
\begin{equation*}
C(\epsilon)=\lim_{N\rightarrow\infty}\frac{1}{N(N-1)}\sum_{i,j=1}^N\theta(\epsilon-\lvert
x_i-x_j\rvert),
\end{equation*}
where $\theta$ is the Heaviside step function
\begin{equation*}
\theta(x)=
\begin{cases}
0,\quad  x<0,\\
1,\quad  x\ge0.
\end{cases}
\end{equation*}
According to Grassberger-Procaccia algorithm, if one knows
$C(\epsilon)$, the fractal dimension of the strange attractor may
be defined
\begin{equation*}
D=\lim_{\epsilon\rightarrow0}\frac{C(\epsilon)}{\log(\epsilon)}.
\end{equation*}
One should calculate $C(\epsilon)$ for different values of
$\epsilon$ and represent the results in coordinates
$\log(\epsilon)$ and $\log(C(\epsilon))$.

The expected dependence of $C(\epsilon)$ has a form of
$\epsilon^D$, so that the obtained graph must have a straight-line
form with $D$ angular coefficient. Results of the numerical
calculations are presented in Fig.4 and Fig.5.
\begin{figure}[tbp]
\centering
\includegraphics[width=7.5cm]{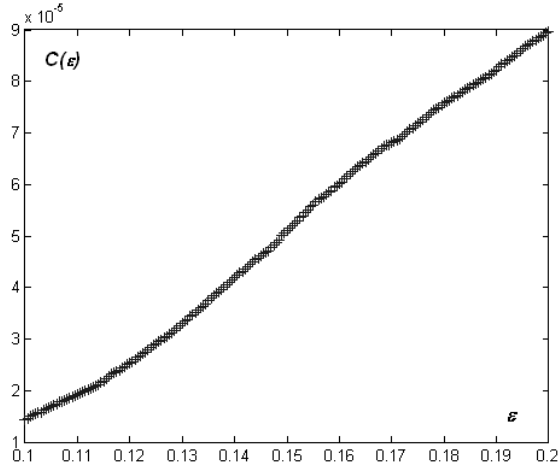}
\caption{\small Graph of the dependence of $C(\epsilon)$ on
$\epsilon$. The graph is plotted via the integration of classical
Hamilton equations corresponding to the Hamiltonian (2.2a) for the
values of parameters $E=-0.125\sim10^{-19} J$, $\lambda=1.0$.}
\label{Fig:4}
\end{figure}
\begin{figure}[tbp]
\centering
\includegraphics[width=7.5cm]{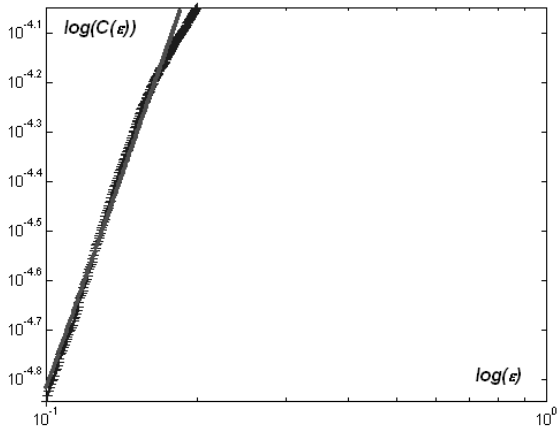}
\caption{\small Graph of the dependence of $C(\epsilon)$ on
$\epsilon$, plotted in a logarithmic scale using the
Grassberger-Procaccia algorithm \cite{17,18,19}. The graph is
plotted via the integration of classical Hamilton equations
corresponding to the Hamiltonian (2.2) for the values of
parameters $E=-0.125\sim10^{-19} J$, $\lambda=1.0$. The curve line
corresponds to the results of the numeric data. The straight-line
is plotted via the processing of the numerical data with least
squares algorithm. From this graph one can estimate the fractal
dimension of the attractor using the formula
$D=\frac{\log(C(\epsilon_2))-\log(C(\epsilon_1))}{\log(\epsilon_2)-\log(\epsilon_1)}\approx2.857.$
}\label{Fig:5}
\end{figure}

\section{Quantum Consideration}
        As it has been already mentioned, usually the quantum chaos is considered as a set of phenomena that appear in
        quantum systems, the classical analogues of which display chaos. The difficulty in definition of quantum chaos is
        connected with a fact, that the concepts characterizing classical chaos (local instability of phase trajectories)
        lose their sense in quantum consideration. Therefore, for describing quantum chaos one has to use methods and
        concepts which are typical for quantum mechanics.

             Our goal in case of quantum consideration, is the demonstration of the fact that one of the signs of the quantum
        chaos is a formation of the mixed state. For endorsement of this
        assumption we shall use methods of random matrix theory \cite{1,28,29}. Due to this fact, we have to use basis of
        eigenfunctions of the integrable part of the Hamiltonian (2.1) and we cannot expend the wave function in
        Landau states, as was done in \cite{15}. This makes situation more complicated but as we are interested in the state $l=0, m=0$ ,
         some analytical results still can be obtained.

             Let us write down the Schr\"odinger equation for the Hamiltonian
             (2.1)
             \begin{equation}
             \hat{H}(\vec{p},\vec{r},\lambda)|\psi(\vec{r},\lambda)\rangle=E(\lambda)|\psi(\vec{r},\lambda)\rangle.
             \end{equation}
             It is easily seen from (3.1), that the Schr\"odinger equation in contrast to the classical equations of Hamilton
             is linear. Like in the case of classical systems, this conclusion means that Eq.(3.1) is completely deterministic
             and does not contain the properties of chaos, the main sign of which is fortuity and abruptness of the
             behavior of system. Because of this the question appears: Then how can one observe nonregular behavior and
             quantum chaos in the quantum systems? To answer this question we shall use some well-known facts from quantum
             mechanics [23,24,25].

                  Let us define via
         $|\psi_{n}(\vec{r},\lambda)\rangle$ the required eigenfunction
        of the Eq.(3.1) and expand it in eigenfunctions of the
        Hamiltonian $\hat{H}(\vec{p},\vec{r},0)=\hat{H}_{0}(\vec{p},\vec{r})$
        \begin{equation}
        |\psi_{n}(\vec{r},\lambda)\rangle=\sum_{m}C_{nm}(\lambda)|\Phi_{m}(\vec{r})\rangle.
        \end{equation}
        Here $|\Phi_{m}(\vec{r})\rangle$  is the eigenfunction of the Hamiltonian
        $\hat{H}_{0}(\vec{p},\vec{r})$.

                      According to the universally recognized rules of quantum mechanics, the coefficients of the
                      expansion $C_{mn}(\lambda)$ taken squared by module, define the probability to find the
                      considered system in state described by
                      wave-function $|\Phi_{m}(\vec{r})\rangle,$  and that is why they satisfy the conditions of normalization:
                      \begin{equation}
                      \sum_{m}|C_{mn}(\lambda)|^{2}=1.
                      \end{equation}
                      Substituting the expansion (3.2) into Schr\"odinger equation (3.1), one can obtain a matrix
                      equation for $C_{mn}$
                      \begin{equation}
                      \sum_{p}H_{kp}C_{pn}=E_{n}C_{kn}(\lambda),
                      \end{equation}
                      where the elements of matrix $H_{kp}$ are defined by the integrals of the type:
                         \begin{equation}
                       H_{kp}=\langle\Phi_{k}(\vec{r})|\hat{H}(\vec{p},\lambda)|\Phi_{p}(\vec{r})\rangle.
                         \end{equation}
                      It is shown in quantum mechanics [23-25] that the series (3.2) exactly converges to the function
                      $|\psi_{n}(\vec{r},\lambda)\rangle$, if when taking sum
                      by $m$  a totality (infinite number) of functions $ |\Phi_{m}(\vec{r})\rangle$ is
                      considered. However the wave function
                      $|\psi_{n}(\vec{r},\lambda)\rangle$ from a qualitative point of view may be with a high degree
                      of accuracy described by the expansion like (3.2) using a finite
                      number $N,$ $(N>>1)$ of states $|\Phi_{m}(\vec{r})\rangle.$The further increase of the
                      number of states $N$ is similar to the tendency of the
                      Plank constant to zero $\hbar\rightarrow0$ and leads to the improvement of
                      quantitative fit between classical and quantum chaos \cite{4,5}.

              According to the premises one may say, that the problem of finding eigenfunctions and eigenvalues
of Hamiltonian $\hat{H}(\vec{p},\vec{r},\lambda)$ may be solved by
a numeric diagonalization of the matrix equation (3.4).

              But here a question is born. If the state of the system is described by a finite superposition of regular
 states $|\Phi_{m}(\vec{r})\rangle,$  then the state $|\psi_{n}(\vec{r},\lambda)\rangle$ is regular too,  so where should
 one look for a nonregular behavior of the system?

         By the assumption taken for today \cite{1,28,29} the properties of chaos in quantum system are revealed
in that sense that matrix (3.5) is a random matrix, all the
elements of which are random values. Based on this, there must
exist some correspondence between matrix (3.5) and the quantities
describing classical chaos. Using our problem as a pattern we
shall start to examine this question.

       Let us start with quantum mechanical investigation of the Hamiltonian (2.2).
 As it has been already mentioned, when the external magnetic field is
 absent$\lambda=0$ the Hamiltonian $\hat{H}(\vec{p},\vec{r},0)$ is exactly integrable.
 When the external magnetic field is turned on $\lambda>1$ the Hamiltonian becomes nonintegrable,
 and for some definite value $\lambda=\lambda_{0}$ the solutions of the classical equations
 corresponding to the Hamiltonian
 $\hat{H}(\vec{p},\vec{r},\lambda_{0})$ display chaos.
 By analogy with classical chaos, when a set of phase trajectories with a small
 dispersion of initial conditions is investigated, we shall consider a small increment of the parameter
 $\delta\lambda=\lambda-\lambda_{0}.$Using the basis of functions in which
 Hamiltonian $\hat{H}(\vec{p},\vec{r},\lambda_{0}) $ is diagonal, for small values of
 $\delta\lambda$ the Hamiltonian $\hat{H}(\vec{p},\vec{r},\lambda)
 $ may be represented as a sum of two matrices:
 \begin{equation}
 \hat{H}=E_{0}+\delta\lambda B,
 \end{equation}
 where $E_{0} $ is a diagonal matrix, $B$ is a banded matrix,
 the elements of which are random numbers. By a numerical diagonalization,
 one can define the eigenfunctions $|\psi_{n}(\vec{r},\lambda+\lambda_{0})\rangle,$
$|\psi_{n}(\vec{r},\lambda)\rangle,$ and eigenvalues
$E_{n}(\lambda_{0}+\delta\lambda_{0}),$ $E_{n}(\lambda_{0})$
corresponding to the $\hat{H}(\vec{p},\vec{r},\lambda)$ and
$\hat{H}(\vec{p},\vec{r},\lambda_{0})$ respectively.
   We are interested in the interval of energy
 $\delta E=E(\lambda_0+\delta\lambda_0)-E(\lambda_0)$
 which is small from a classical point of view, and large from a quantum mechanical one
 (with relation to the large number of quantum levels contained in this interval).

     Let us begin to the quantum mechanical analysis of the Hamiltonian (2.1).
 The eigenvectors of the basis, in which Hamiltonian  is diagonal, can be found easily
 and are of the form \cite{23}
 \begin{equation}
 |\Phi_{n,l,m}(r,\theta,\varphi)\rangle=|R_{n,l}Y_{lm}(\theta,\varphi)\rangle,
 \end{equation}
 where $R_{nl}=\frac{2}{n^{l+2}(2l+1)!}\sqrt{\frac{(n+l)!}{(n-l-1)!}}(2r)^{l}e^{-r/l}F(-n+l+1,2l+2,\frac{2r}{n})$
  is the normalized radial part of the wave function and $Y_{lm}(\theta,\varphi),$
  $F(-n+l+1,2l+2,\frac{2r}{n})$ are the spherical and hyper-geometrical
  functions respectively \cite{27};  $r,$ $\theta$ and $\varphi$  denotes the spherical coordinates.

          As was mentioned above, using of the symmetric Kepler problem's eigenfunctions is connected
 to the requirements of the random matrix theory. Calculation of non-diagonal matrix elements in
 this basis is more complicated then in Landau basis \cite{15,23}. But the fact that we are interested in the
 state $l=0,$ $m=0$ makes life and it's still possible to obtain some analytical results. Actually the problem is reduced to the
 calculation of the following matrix elements:
 \begin{equation}
 \langle\Phi_{n,0,0}(r,\theta,\varphi)|x^{2}+y^{2}|\Phi_{m,0,0}\rangle=
 \langle\Phi_{n,0,0}(r,\theta,\varphi)|r^{2}sin^{2}\theta|\Phi_{m,0,0}(r,\theta,\varphi)\rangle.
 \end{equation}

 These matrix elements can be calculated analytically.
 As a result for diagonal matrix elements we get \cite{25}:
 \begin{equation}
 \langle\Phi_{n,0,0}(r,\theta,\varphi)|r^{2}sin^{2}\theta|\Phi_{n,0,0}\rangle=\frac{n^{2}}{2}(5n^{2}+1)
\end{equation}
and non-diagonal matrix elements can be reduced to the calculation
of the following integral:
\begin{equation}
\langle\Phi_{n,0,0}(r,\theta,\varphi)|r^{2}sin^{2}\theta|\Phi_{m,0,0}(r,\theta,\varphi)\rangle=\frac{1}{3n^{3/2}m^{3/2}}
J^{S,0}_{\gamma}(\alpha,\alpha^{'}),
\end{equation}
$$J^{S,0}_{\gamma}(\alpha,\alpha^{'})=
\int_{0}^{\infty}e^{-\frac{k+k^{'}}{2}}F(\alpha,\gamma,kr)F(\alpha^{'},\gamma,k^{'}r)dr,$$
where
$$ k=\frac{2}{n};k^{'}=\frac{2}{m};\alpha=-n+1;\alpha^{'}=-m+1;\gamma=2;S=3.$$
Direct calculation of the integral
$J_{\gamma}^{S,0}(\alpha,\alpha^{'})$  is not possible. But we can
use recurrence formula, which connects integral
$J_{\gamma}^{S,0}(\alpha,\alpha^{'})$ and integrals with the lower
values of superscript $J_{\gamma}^{S-1,0}(\alpha,\alpha^{'})$;
$J_{\gamma}^{S-2,0}(\alpha,\alpha^{'})$. After straightforward but
laborious calculations one can
obtain $J_{\gamma}^{S,0}(\alpha,\alpha^{'})$:
$$ J^{S,0}_{\gamma}=\frac{4}{k^{2}-k^{'2}}[(\frac{\gamma}{2}(k-k^{'})-k\alpha+k^{'}\alpha^{'}-k^{'}(S-1))
J^{S-1,0}_{\gamma}(\alpha,\alpha^{'})+$$
$$+(S-1)(\gamma-1+S-1-2\alpha^{'})J_{\gamma}^{S-2,0}(\alpha,\alpha^{'})+2\alpha^{'}(S-1)J_{\gamma}^{S-2,0}(\alpha,\alpha{'}+1)]$$
where
\begin{equation}
J^{0,0}_{\gamma}(\alpha,\alpha^{'})=2^{\gamma}\Gamma(\gamma)(k+k^{'})^{\alpha+\alpha^{'}-\gamma}
(k^{'}-k)^{-\alpha}(k-k^{'})^{-\alpha^{'}}F(\alpha,\alpha^{'},\gamma,-\frac{4kk^{'}}{(k-k^{'})^{2}})
 \end{equation}
and $\Gamma(\gamma)$ is the Euler gamma function \cite{27}.

\section{Irreversible evolution of the quantum chaos. Formation of the mixed state.}

      At the discussion of the nonreversible evolution of the
 quantum -mechanical system, naturally the question emerges.
 How in a quantum system originates non -reversibility? If the
 quantum system evolves according to the Schr\"odinger equation,
 how can a pure quantum -mechanical state become a  mixed one? The
 question is that, in contrast to the classical chaos, quantum
 mechanically irregular motion cannot be characterized by extreme
 sensitivity to tiny changes of the initial data. Due to
 the unitarity of the quantum dynamics, the overlap of two wave
 functions remains time-independent $|\langle\vartheta(t)|\zeta(t)\rangle|^{2}=|\langle\vartheta(0)|\zeta(0)\rangle|^{2},$
 provided time-dependence of $\vartheta(t)$ and $\zeta(t)$ is
 generated by the same Hamiltonian. However, an alternative
 characterization of classical chaos, extreme sensitivity to
 slight changes  of the dynamics does carry over into quantum
 mechanics.

    Let us assume, that the amplitude of the external magnetic
 field is modulated by varying field
\begin{equation}
B=B_{0}+\Delta B_0cos\Omega t.
\end{equation}
Taking into account (4.1) the Schr\"odinger equation for the wave
function of diamagnetic hydrogen atom takes the form
\begin{equation}
i\hbar\frac{\partial|\psi(\vec{r},\lambda(t))\rangle}{\partial
t}=\hat{H}(t)|\psi(\vec{r},\lambda(t))\rangle,
\end{equation}
where
$$ \hat{H}(t)=\hat{H}_{0}(\vec{p},\vec{r},\lambda_{0})+\Delta\lambda(t)V(r,\theta),$$
$$\Delta\lambda(t)=\Delta\lambda cos\Omega t.$$

    The solution of the time-dependent Schr\"odinger equation
 formally can be written with the help of a
 time-ordered exponential

 $$
 U(t)=\{exp[-\frac{i}{\hbar}\int_{0}^{t}dt^{'}\hat{H}(t^{'})]\}_{+},$$
  where the positive time ordering requires
  $$ [A(t)B(t^{'})]_{+}= \left\{ \begin{array}{ll}
                            A(t)B(t^{'}) & \mbox{if $t>t^{'}$} \\B(t^{'})A(t) & \mbox{if $t<t^{'}$} \end {array}\right.$$
  In our case  $H(t+\frac{2\pi k}{\Omega})=H(t),$ the evolution
  operator referring to one period $T_{0}=\frac{2\pi}{\Omega},$
  the so-called Flouqet operator $U(T_{0})\equiv\hat{F}$ \cite{1},
  is worthy of consideration since it yields a stroboscopic view
  of the dynamics
  \begin{equation}
  \psi(kT_{0})=(\hat{F})^{n}\psi(0).
  \end{equation}
   The Floquet operator, being unitary, has unimodular
   eigenvalues. Suppose we can find eigenvectors
   $|\varphi_{\chi}\rangle$ of the Floquet operator
   \begin{eqnarray}
   \hat{F}|\varphi_{\chi}\rangle&=&e^{-i\varphi_{\chi}}|\varphi_{\chi}\rangle,\\
   \langle\varphi_{\chi}|\varphi_{\theta\rangle}&=&\delta_{\chi\theta}.
   \nonumber
   \end{eqnarray}
 Then, with the eigenvalue problem solved, the stroboscopic dynamics can be written out
 explicitly \cite{1}
\begin{equation}
\psi(kT_{0})=\sum_{\chi}e^{-ik\varphi_{\chi}}\langle\varphi_{\chi}|\psi(0)\rangle|\varphi_{\chi}\rangle.
\end{equation}
As it was mentioned above, our aim is to proof that one of the
signs of the emergence of quantum chaos is a formation of the
mixed state. Being initially in a pure quantum mechanical state,
described by the wave function $\psi_{n},$ the system during the
evolution makes an irreversible transition to the mixed state.

After the formation of a mixed state the quantum-mechanical
description loses its sense and we have to use a
quantum-statistical interpretation. To prove the formation of the
mixed state, one has to show the zeroing of non-diagonal elements
of density matrix, the equality of which to zero is a sign of a
mixed state \cite{30,31}.

Taking (4.5) into account, we obtain for non diagonal matrix
elements of the density matrix:
\begin{equation}
\rho_{nm}=\overline{C_nC^*_m},
\end{equation}
where $C_n=e^{-ik\phi_n}\langle\phi_n|\psi(0)\rangle$,
$\overline{(...)}$ means averaging over time.

For averaging let us recollect that value $\phi_n$ is the
eigenvalue of the Floquet operator. Owing to the non-integrability
of the system, eigenvalues can be obtained only by way of
numerical diagonalization of the Hamiltonian $\hat{H}$ (4.2).
According to the main hypothesis of the random matrix theory
\cite{1,11}, elements of this matrix in the chaotic domain are
random values. So, their eigenvalues also can be considered as
random values. This statement is valid only in the chaotic domain
\cite{32}. According to this, the phase
\begin{equation}
f(n,m)=\phi_m-\phi_n,
\end{equation}
also may be considered as random. Therefore when $T=kT_0$, $k\gg
1$, we get
\begin{equation}
\rho_{nm}=\frac{T_0}{T}\sum^{T/T_0}_{k=1}\langle
e^{-ikf(n,m)}\rangle\langle\phi_m|\psi(0)\rangle\langle\psi(0)|\phi_n\rangle=0,
\end{equation}
where $\langle ...\rangle$ denotes ensemble average of random
matrices, that corresponds to a small dispersion of the value of
magnetic field parameter  $\delta\lambda_0$.

Expression (4.8) may be found by more vigorous mathematical
substantiation. For this let us recall some details from the
probability theory \cite{33}.

a)  In general case, under the characteristic function of the
random value $X$, mathematical expectation of the following
exponent is meant
\begin{equation}
F(t)=M(\exp(itX)),
\end{equation}
where $t$ is a real parameter.

b)  Mathematical expectation itself is defined as a first moment
$\mu_1$ of the random variable $X$
\begin{equation}
M(X)=\langle X\rangle=\mu_1=\sum_k x_kP_k,
\end{equation}
where $X$ is the discrete random variable which takes possible
values $x_1, x_2, ...$ with appropriate probabilities
$P_k=P(X=x_k)$, $\langle ...\rangle$ means average.

Taking (4.9), (4.10) into account and considering $f(n,m)$ as a
random value, we get:
\begin{equation}
\rho_{nm}=\frac{T_0}{T}\sum^{T/T_0}_{k=1}F(k),
\end{equation}
where
\begin{equation}
F(k)=M(e^{ikf(n,m)}),
\end{equation}
is the mathematical expectation of the characteristic function of
the random phase $f(n,m)$, and
\begin{equation}
\nonumber
A=\langle\phi_m|\psi(0)\rangle\langle\psi(0)|\phi_n\rangle.
\end{equation}
Let us assume that random phase $f(n,m)$ has a normal dispersion
\cite{33}. Then from (4.12) one can obtain
\begin{equation}
F(k)=e^{iak}e^{-\frac{\sigma^2k^2}{2}},
\end{equation}
where
\begin{equation}
a=M(f(n,m)),
\end{equation}
is the mathematical expectation of the random phase,
\begin{equation}
\sigma^2=M(f^2(n,m))-(M(f(n,m)))^2,
\end{equation}
is the mean square deviation.

Substituting (4.13) in (4.11) and making transition from summation
to the integration $\sum_k \rightarrow \int dk$ finally we get
\begin{equation}
\rho_{nm}(T)=\frac{C_0}{T}(Erfi[\frac{a}{\sqrt{2}\sigma}]+Erfi[\frac{a+i(T/T_0)\sigma^2}{\sigma}]),
\end{equation}
where $C_0=iAe^{\frac{a^2}{2\sigma^2}}$ and $Erfi(...)$ is the
error function, connected to the ordinary error function \cite{33}
by the relation
\begin{equation}
Erfi(x)=-i\frac{\sqrt{\pi}}{2}Erfi(ix).
\end{equation}
Taking into consideration asymptotical form of the function
$Re(Erfi(x)) \rightarrow 0$, $x \rightarrow \infty$ and
$Im(Erfi(x)) \rightarrow 1$, $x \rightarrow \infty$, in case when
distribution of the eigenvalues of the quantum-mechanical chaotic
system obeys the normal distribution, we have
\begin{equation}
\begin{split}
|\rho_{nm}(T)| \sim 1/T \sim 0, \\
T \gg T_0.
\end{split}
\end{equation}

 \begin{figure}[h]
 \centering
 \includegraphics[width=7.5cm]{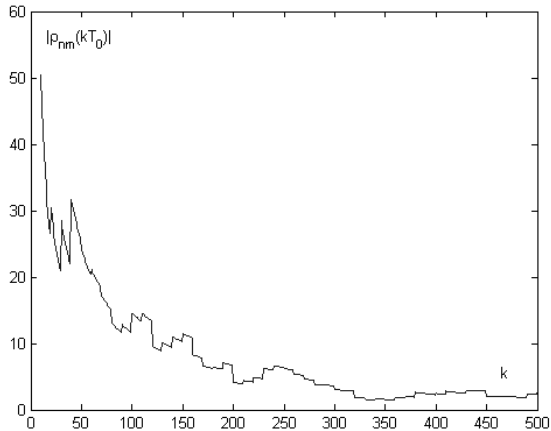}
 \caption{\small The graph of the dependence of the second term  $\rho_{mn}$ of
 non-diagonal matrix element (4.6) on time
 $t=kT_0$. The graph is obtained using (4.8) and by way of
 numerical diagonalization of the Hamiltonian (4.2) for different
 values $\hat{H}(t_i)=\hat{H}(\lambda(t_i))$, $t_i\in[0,T]$. The
 graph is plotted after taking an ensemble average of Hamiltonians
 $H(\lambda_0)$, corresponding to the dispersion $\delta\lambda_0=0.1$ of the value of
 magnetic field  $\lambda_0$.
 As it is easily seen from the graph that when
 $t=kT_0>\tau_c$, $T_0=\frac{2\pi}{\Omega}=0.1\tau_c$, zeroing of
 $\rho_{nm}$ happens. In the numeric calculations we have
 used definite initial conditions $\psi_n(0)$, $\psi_m(0)$,
 obtained via numerical diagonalization of the Hamiltonian (2.1)
 for $\lambda_0=1$. The dimension of the diagonalized matrix was
 $500\times 500$, $\chi,\zeta=\overline{1,500}$}.  \label{Fig:6}
 \end{figure}

     Zeroing of the non-diagonal elements of density matrix is the sign of quantum chaos beginnings and
formation of mixed state.  From that moment the quantum dynamics is
non-reversible since the information about the wave functions
phase is lost. Given above reasoning can be checked by numerical
experiment, which consists in: the direct solving of
non-stationary Schr\"odinger equation (4.2), defining eigenvalues $\varphi_\chi$ and in estimation of the
averaged non-diagonal elements of the density matrix according to (4.8).

The result of numerical calculations is represented on Fig.6.

\section{Quantum Poincar\'e Recurrences}

According to the results of the previous section (see Fig.6), for times $t>\tau_c$ formation of a mixed state happens
due to time evolution of quantum-mechanical system. Consequently the concept of a wave function loses its meaning and the
Poincar\'e recurrences are out of the question. But what happens when $t<\tau_c$? We shall clear up this question in the
given section.

As it was mentioned above, in work \cite{22} the concept of
quantum Poincar\'e recurrence was introduced. In particular, the
authors have studied the quantum dynamics of one-dimensional
rotator under the influence of delta-like perturbation. The
definition of the "distance" between the wave-functions was
introduced
\begin{equation}
d^2(t)=\langle\Psi(\vec{r},t)-\Psi(\vec{r},0)\vert\Psi(\vec{r},t)-\Psi(\vec{r},0)\rangle,
\end{equation}
where $\vert\Psi(\vec{r},t)\rangle$ is the solution of nonstationary
Schr\"odinger equation
\begin{equation}
i\hbar\frac{\partial\Psi}{\partial t}=\hat{H}(t)\Psi.
\end{equation}
The authors of the work, cited above, look for the solution of the
equation (5.2) in the form of expansion of the function
$\vert\Psi(\vec{r},t)\rangle$ in an arbitrary orthogonal
basis $\vert\Phi_m\rangle$
\begin{equation}
\vert\Psi(\vec{r},t)\rangle_n=\sum_mC_{nm}(t)\vert\Phi_m(\vec{r})\rangle,
\end{equation}
where $C_{nm}(t)$ are the expansion coefficients, which determine
the time dependence of the function
$\vert\Psi(\vec{r},t)\rangle_n$. It is obvious, that
during the evolution in time of the wave-function
$\vert\Psi(\vec{r},t)\rangle$, the quantity $d(t)$ will
change in time. The authors offer to consider the quantum
Poincar\'e recurrences as such values of time $t$, that satisfy
the condition $\lvert d(t)\rvert<\epsilon$, where $\epsilon$ is a
small constant value ($\epsilon<1$).

\begin{figure}[h]
\centering
\includegraphics[width=7.5cm]{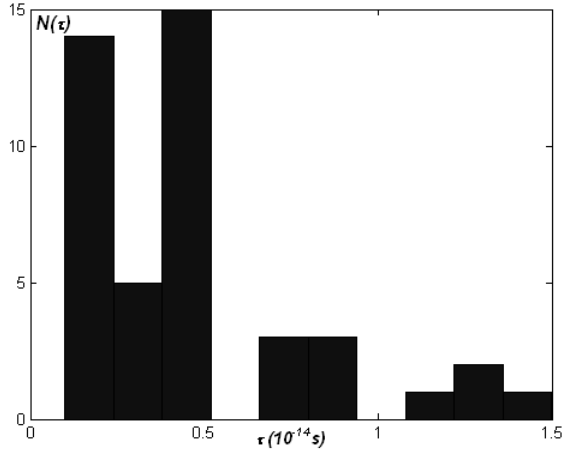}
\caption{\small The histogram of distributions of the times of
quantum Poincar\'e recurrences, plotted in compliance with the
definition (5.1). The quantity $N(\tau)$ determines the dependence
of  the number of quantum Poincar\'e recurrences on the
distribution of the periods of recurrences.$\lambda_0=1$,
$\Delta\lambda=0.1$, $\Omega=\frac{2\pi}{T_0}$,
$T_0\sim0.1\tau_c\approx0.5\cdot10^{-15} s$.} \label{Fig:7}
\end{figure}

     Our purpose is the use of this method for the study of quantum
Poincar\'e recurrences in the system of diamagnetic hydrogen atom.
For that, like in the previous section, we shall assume, that the amplitude of the external
magnetic field is modulated by a varying field. But in contrast to the previous section, we
shall examine the solution of the equation in interval $t<\tau_c$, when the system is still in a pure quantum-mechanical state
and chaotization of phase and formation of a mixed state can be neglected.

   Before starting the analysis of the equation (4.2), let us note
some typical differences of our problem from the problem studied
in \cite{22}. In problem studied in \cite{22}, the dimension of
the Hilbert space is finite. That is why the wave-function
$\vert\Psi(\vec{r},t)\rangle_n$, according to (5.3), is
exactly defined by the finite set of the orthogonal functions
$\vert\Phi_m(\vec{r})\rangle$. The authors make an assumption,
that the expansion coefficients $C_{nm}(t)$ are the random
quantities and satisfy the Gaussian distribution.

In our case the dimension of the Hilbert space is infinite and
the approximation of $\vert\Psi(\vec{r},t)\rangle_n$ by
the finite set of functions of the orthogonal basis
$\vert\Phi_m(\vec{r})\rangle$ is rough. Our purpose is a proof
that in spite of it the quantum Poincar\'e recurrences may happen
in the system.

\begin{figure}[h]
\centering
\includegraphics[width=7.5cm]{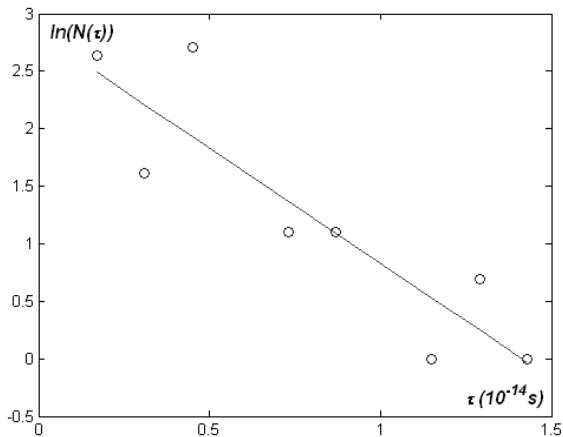}
\caption{\small Exponential approximation of the numeric data
represented in Fig.7, made in compliance with the formula (2.4). The graph
is plotted in a logarithmic scale. Therefore a straight line on the given graph
corresponds to the exponential Lorentz form.
The result allows to determine numerically the mean time of
quantum Poincar\'e recurrences $\tau_{rec}^Q$ and to compare it
with the mean time of classical Poincar\'e recurrences
$\tau_{rec}\sim\tau\sim 1/h_0$ (see formula (2.4)). As it comes
from the graph, $\tau_{rec}^Q\sim\frac{1}{\lvert
b\rvert}\approx0.5\cdot10^{-14} s$, where $b$ is the angular
coefficient of the approximating straight-line. The achieved
estimate for the mean time of quantum recurrences is in a good
agreement with the mean value of the classical Poincar\'e
recurrences
$\tau_{rec}\sim\tau_c\sim\frac{1}{\delta\omega}=0.5\cdot10^{-14}
s$ (see Fig.3).} \label{Fig:8}
\end{figure}

For the solution of the nonstationary Schr\"odinger equation (4.2)
we shall consider $t$ as a discrete quantity (consequently the
values of the parameter $\delta\lambda(t)$ become discrete too).
We shall not assume the coefficients $C_{nm}(\lambda,t)$ to be
\textit{a priori} random, but we shall determine their values by
solving the matrix equation (3.5) and taking into account the
dispersion of the parameter $\delta\lambda_0$. Thus, using the
formulas (3.2), (3.5) and (5.1) one can define the distribution of
the times of Poincar\'e recurrences. The results of numeric
calculations are shown in Fig.7 and Fig.8.

According to Fig.8, the distribution of the times of Poincar\'e
recurrences is of the Lorenz form. This fact can be easily
explained, taking into account the existence of a strong classical
chaos (i.e. the absence of the stability islands in the phase
space).

The Lorentz form testifies that if $t>\tau_{rec}^Q=\tau_c$, quantum Poinacr\'e
recurrences are improbable. This coincides with the results of the previous section,
according to which the system is found in a mixed state in this time interval.

Based on the results obtained above, one may assert that quantum
Poincar\'e recurrences also take place in quantum systems with
infinite "phase space" (i.e. with infinite-dimensional Hilbert
space).

\end{document}